# Computer Simulation-Based Learning: Student Self-Efficacy During COVID-19 Outbreak

Thaweesak Trongtirakul
Dept. of Electrical Engineering
Rajamangala University of
Technology Phra Nakhon
Bangkok, Thailand
thaweesak.tr@rmutp.ac.th

Kamonnit Pusorn
Dept. of Electrical Engineering
Rajamangala University of
Technology Phra Nakhon
Bangkok, Thailand
wanida.pu@rmutp.ac.th

Umpaporn Peerawanichkul
Dept. of Computer Engineering
Rajamangala University of
Technology Phra Nakhon
Bangkok, Thailand
umpaporn.p@rmutp.ac.th

*Abstract*—Due to the COVID-19 as a pandemic, the government has forced the nationwide shutdown of several activities, including educational activities. It has resulted in gigantic migration of universities with education over the internet serving as the educational platform. Hand-on-based learning becomes a new challenge. This paper aims to investigate the effect of computer simulation-based learning on student self-efficacy in an electric circuit analysis course. For the 17 participants included in this study, the students have overcome their existing achievements indicated by a long-term average score. Computer simulation-based learning provides positive results on student self-efficacy. Students also perceived a valuable learning experience.

*Keywords*—Computer Simulation-Based Learning, Electric Circuit Analysis, Self-Efficacy Assessment

## I. Introduction

Sciences, Technology, Engineering, and Mathematics (STEM) contain complex models of real-world systems. STEM students need to have special training to make decisions that affect complex systems with enormous variables and interactions among their components [1]. In addition, real-world systems are usually subject to uncertainty, which is generated by human configuration-based systems [2-4]. These facts cannot be completely included in conventional analytical models, and require the use of computer simulation-based methods in coordination with other techniques, such as nature-inspired optimizations [5-7] and supervised models [8-12]. The use of computer simulation programs, tools, and games encourages the theoretical and practical understanding of these complex systems and allows students to improve their learning capability through the development of hands-on activities appropriately designed by their instructors [13-16].

Learning practices by Computer Simulation-based Learning (CSL) programs, tools, and games are interesting in academic and industrial societies, who consider in CSL the way to develop their students and future employees. Figure 1 illustrates the number of articles in Google Scholar searched by the term "Computer Simulation-Based Learning" from 2000 to 2021. It notes that the number of research articles has been published in 2000 with 32,200 articles to more than 123,000 articles in 2021. Also, regression methods have been applied to forecast for the next few years. Based on the statistical forecast, the number of articles will reach the level of 140,000 articles in 2025, approximately.

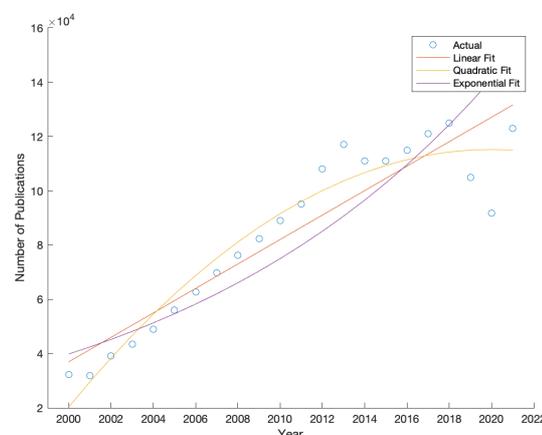

Fig. 1. The number of articles in Google Scholar searched by the term "Computer Simulation-Based Learning"

Engineering education relies on the availability of different laboratories; however, the scenarios universities are exposed to vary depending on the financial subsidization by a government. Individual learning experiences vary in students' educational backgrounds, resulting in possibly missed learning opportunities.

To address the limitation of laboratory-based education, CSL has been proposed as an additional educational strategy. CSL demonstrates to recreate real-world experiments through simulated scenarios, environments, providing a safe environment [17-20]. With CSL, learning experiences can be customized to specific learning and can be set up on-demand, eliminating the lack of laboratory availability.

This paper aims to investigate the effect of CSL on student self-efficacy in the electric circuit analysis assessment. The remaining sections of the paper are structured as follows. In section II, the electric circuit simulation with MATLAB® Simulink is presented. Section III exhibits the results and discussion of the proposed CSL. Finally, the paper ends with the conclusion in Section IV.

## II. ELECTRIC CIRCUIT SIMULATION WITH MATLAB® SIMULINK

This study was a prospective, experimental study using weekly scores to assess student performance with CSL as a learning strategy.

### A. Participants

Students studying in the electrical engineering major at Rajamnagala University of Technology Phra Nakhon (RMUTP), Faculty of Industrial Education, enrolled in the Electric Circuit Analysis academic course of study, which was delivered in the 1st year of their electrical engineering program.

### B. Outcome Measures

The primary outcome measured in this research project was student self-efficacy in the electric circuit analysis assessment. Student self-efficacy was measured using weekly self-scores calculated by using 4-week moving average scores.

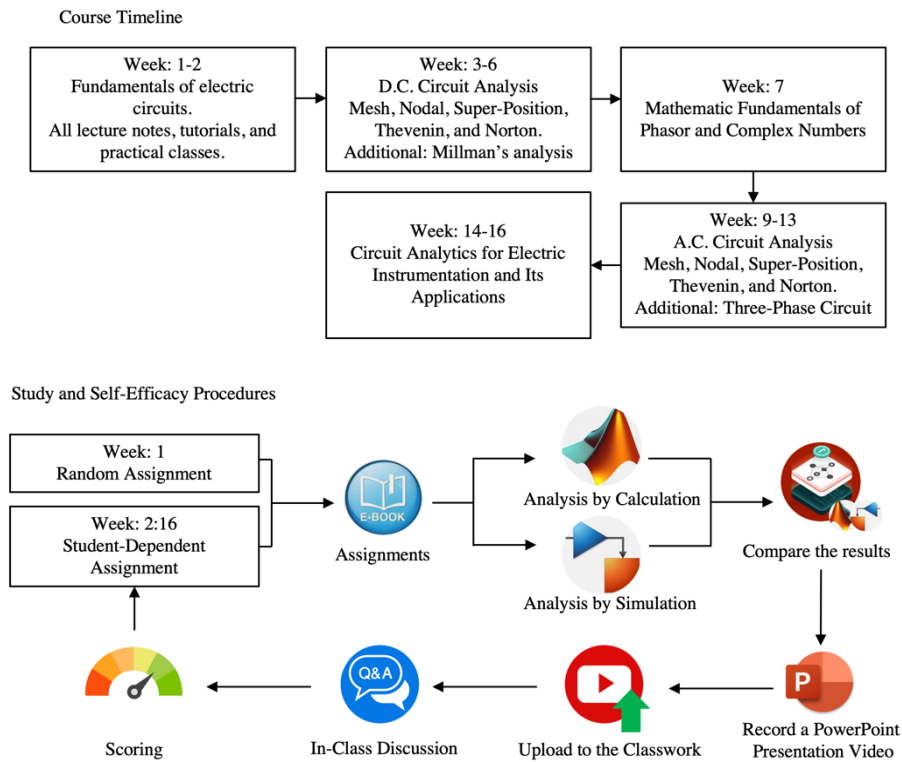

Fig. 2. Timeline of Study Procedure

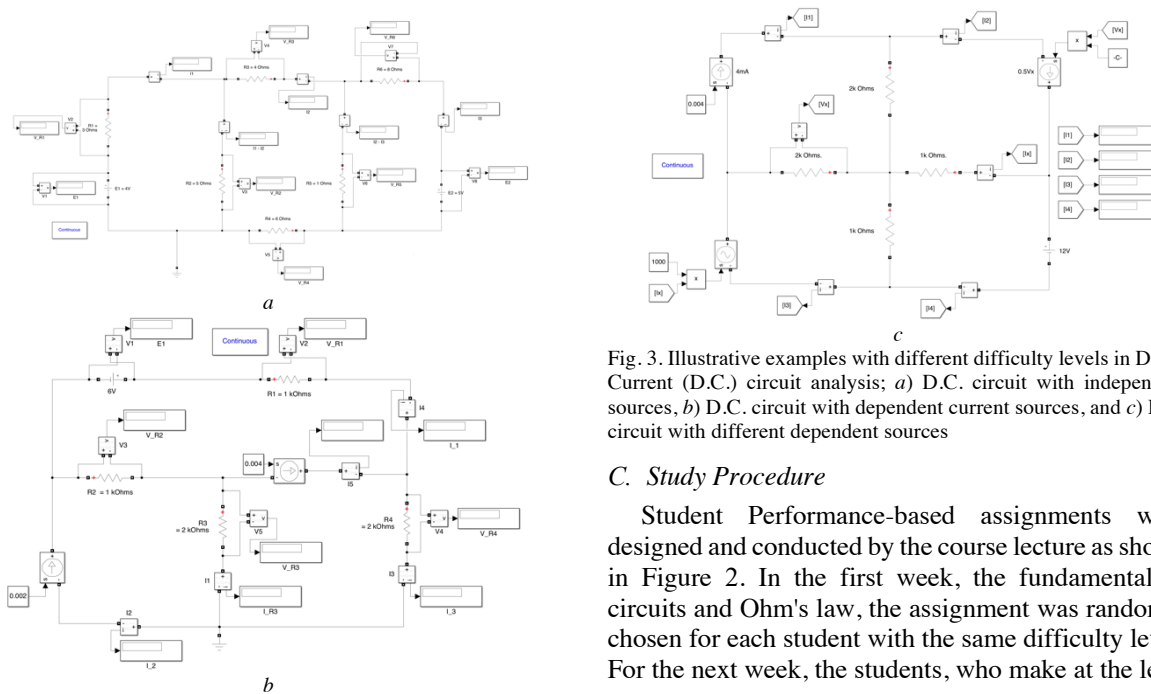

Fig. 3. Illustrative examples with different difficulty levels in Direct Current (D.C.) circuit analysis; *a*) D.C. circuit with independent sources, *b*) D.C. circuit with dependent current sources, and *c*) D.C. circuit with different dependent sources

### C. Study Procedure

Student Performance-based assignments were designed and conducted by the course lecture as shown in Figure 2. In the first week, the fundamental of circuits and Ohm's law, the assignment was randomly chosen for each student with the same difficulty level. For the next week, the students, who make at the least score, could choose the assignments, providing in three difficulty levels: Normal, Difficult, and Very Difficult shown in Figure 3, with different score levels.

The students must upload a recorded PowerPoint presentation video to the Google Classroom, a learning platform that aims to simplify creating, distributing, and scoring assignments. The circuit analytical mistakes of students were discussed with the class each week.

*D. Statistical Analysis*

To analyze the pre-post CSL in terms of self-efficacy for each of the students, weekly scores were calculated for measuring the electric analytical development by using 4-week moving averages. The weekly scores outperforming their moving average scores were considered significant.

### III. RESULTS AND DISCUSSIONS

Based on the cohort of 18 students enrolled in the Electric Circuit Analysis unit in 2021, the student self-efficacy results were analyzed as illustrated in Figure 4. It shows that students were able to score very well during the $1^{st}$ - $3^{rd}$ weeks. The partial content in the beginning course overlaps the content of a high-vocational degree – the previous degree of the student. In the fourth week, the continuous content is more and more difficult. The fundamentals of electric circuit analysis are extended from the previous content. The fourth week's average score (blue line) of students considerably drops. It shows that student self-efficacy underperforms their existing performance. With the content of the videos tailored by a course lecturer, the students attempt to study with MATLAB® – a computer simulation program for engineering. Students have improved their self-efficacy with progressive scores. Since the sixth week, students have overcome their existing and current achievements. Finally, the student self-efficacy outperforms the long-term average of the entire semester.

The analytics with weekly scores divided by the number of video views each week reveals the rate of self-learning through clip videos could improve student self-learning as shown in Figure 5. Interestingly, the number of video views in the 3rd week is lower than expected. Based on our student interviews, we found that most students have excellent capability in circuit analysis by using the Mesh analysis. On the other hand, students have never analyzed circuits with the Nodal method. The number of views soars by 4.3 times of average scores in the next week, but the score of students underperformed. We have discussed the assignments to see weaknesses and strengths. Also, strengths parts have been brought up while weaknesses have been eliminated. It is evident that students could make higher scores with less video attention when compared with the last week and could learn better with less time. It is concluded that our videos could help the student to get more understanding in calculating and analyzing circuits with MATLAB® Simulink.

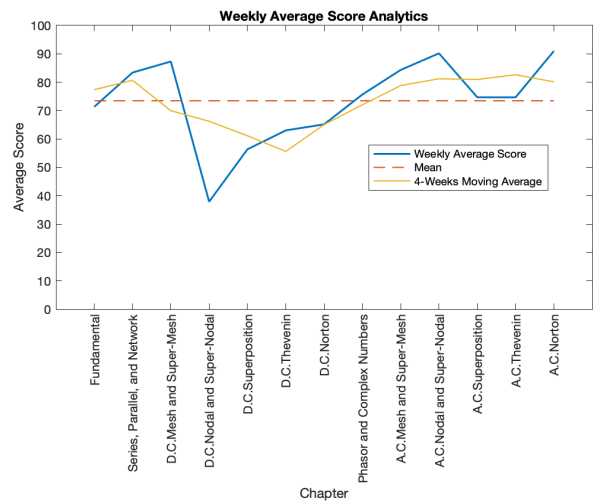

Fig. 4. Student self-efficacy evaluated by weekly average score

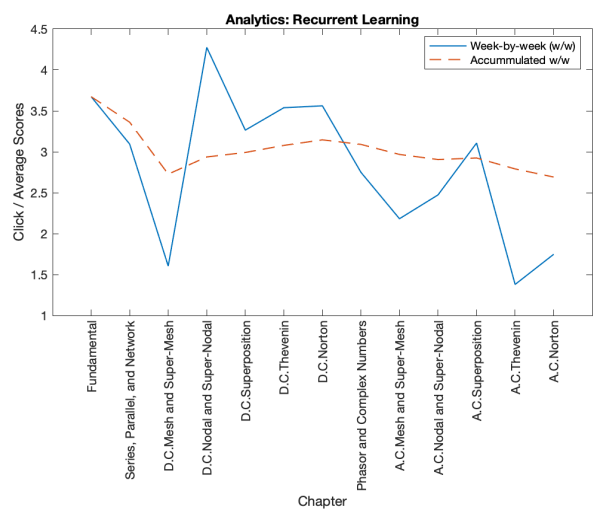

Fig. 5. Relationship between weekly scores and the number of video views.

### IV. CONCLUSION

The results of the self-efficacy were clearly positive with significant improvement to student self-efficacy. CSL can guarantee the provision of consistent and diverse learning experiences and include exposure to scenarios that are engineering unusual, promoting a more equitable learning experience for all students. Students also reported that they appreciated CSL to be a wealth of learning experience.

### ACKNOWLEDGMENT

The authors would like to acknowledge Prof. Sos Agaian for providing his teaching philosophy and the wealth recommendation from his teaching experience. Also, Thanks to Teaching Assistant Krittayot Kuisorn for all of his hard-working, creating an inclusive and supportive learning environment.